\documentclass[12pt]{article}
\usepackage{amssymb}

\title{Critical behavior of the 3-state Potts model on Sierpinski carpet}
\author{ Pai-Yi Hsiao$^{1}$ and Pascal Monceau$^{1,2}$\\
\small
$^{1}$Laboratoire de physique th\'{e}orique de la mati\`{e}re condens\'{e}e,\\
\small
Universit\'e Paris VII -- Denis Diderot, 
case 7020, 2 place Jussieu, 75251 Paris cedex 05, France\\
\small
$^{2}$D\'{e}partement de physique et mod\'{e}lisation, \\
\small
Universit\'{e} d'Evry-Val d'Essonne, Boulevard F. Mitterrand, 
91025 Evry cedex, France
} 
\begin{document}
\maketitle

\begin{abstract}
We study the critical behavior of the 3-state Potts model, where the
spins are located at the centers of the occupied squares of the deterministic 
Sierpinski carpet. 
A finite-size scaling analysis is performed from Monte Carlo simulations, for
a Hausdorff dimension $d_{f}$ $\simeq 1.8928$. 
The phase transition is shown to be a second order one. 
The maxima of the susceptibility of the order parameter follow a power
law in a very reliable way, which enables us to calculate the ratio of the
exponents $\gamma /\nu $. 
We find that the scaling corrections affect the behavior of 
most of the thermodynamical quantities. 
However, the sequence of intersection points extracted from the Binder's 
cumulant provides bounds for the critical temperature. 
We are  able to give the bounds for the exponent $1/\nu$ as well as 
for the ratio of the exponents $\beta/\nu $, which are compatible with the 
results calculated from the hyperscaling relation.
\end{abstract}


\section{Introduction}
\qquad Critical phenomena on fractal structures have been first studied by
Gefen {\it et al.} \cite{gefen80}. 
Since then, most of the works on this topic have been done in the framework 
of the Ising model. 
Recently, the use of cluster algorithms and histogram methods enabled two 
groups \cite{monceau98, carmona98, hsiao00, monceau01} to study the 
ferromagnetic Ising transition on the fractal lattices by Monte Carlo 
simulation much more thoroughly than before; 
several fractal dimensions between $1$ and $3$ have been investigated. 
These last results show that the scaling corrections can strongly
affect the finite size behavior of some thermodynamical quantities,
especially when the fractal dimension is lowered from $2$ to $1$; the
critical exponents cannot always be calculated, but bounds can be provided.
The available values and bounds of the exponents are consistent with the
hyperscaling relation provided that the space dimension is replaced by the
Hausdorff one. Furthermore, they do not agree with $\epsilon $-expansion
results \cite{guillou87}. The critical behavior of fractals is said to be
understood in the framework of weak universality \cite{gefen80, hsiao00,wu87}%
: the critical exponents do not only depend upon the symmetry of the order
parameter, the interaction range, and the fractal dimension, but also upon
geometrical features of the fractal structure.

The Potts model is a generalization of the Ising model obtained by varying
the number of spin states from $2$ to any non zero value $q$. 
Since the number of states $q$ is related to the symmetrical properties of 
the order parameter, the critical behavior of the Potts model depends upon 
the value of this additional variable \cite{wu82}. 
One of the most striking results is the effect of $q$ on the order of the 
transition as a function of the space dimension $d$: In the case of regular 
translational invariant lattices, it has been shown that the ferromagnetic 
phase transition is a second order one if $q$ is smaller than a value 
$q_{c}(d)$ and a first order one if $q>$ $q_{c}(d)$. 
For instance, the phase transition in the bidimensional case is first order 
for $q>4$ and second order for $q\leq 4$;
the value of $q_{c}(3)$ is smaller than $3$. 
Moreover, a wide variety of effects occur in the presence of disorder: 
bond randomness can induce a second order transition from a system exhibiting 
a first order one \cite{chen95}. 
The order of the transition can be changed to a second  one
if strong enough aperiodic fluctuations are introduced \cite{berche98}.

The question arises whether the values $q_{c}(d)$ make sense in non integer
dimensions $d_{f}$ or not. Fractals are natural candidates to interpolate
between integer space dimensions. Since translation invariance is broken in
these structures, disorder is introduced: the lattice is strongly
inhomogeneous. Thus, two among the three parameters driving the critical
behavior of the Potts model (structured fractal disorder and dimensionality) 
are closely linked in the present case of fractals. 
The critical properties of the Potts model on
fractal structures have never been studied before, excepted by Bin 
\cite{bin86} who generalizes the Migdal-Kadanoff \ bond-moving renormalization
scheme on the Sierpinski carpet from the Ising model \cite{gefen84} to the
Potts one. No Monte Carlo results are available, up to now; there are two
main reasons why studies of the critical behaviors on fractal
structures is a difficult task:
\begin{enumerate}
\item[(i)]
Monte Carlo simulations come up against critical slowing down.
Fortunately, the use of cluster algorithms \cite{wolff88, sw87} is
very helpful in overcoming this difficulty.
\item[(ii)]
The scaling corrections are expected to be large \cite{monceau01}. 
A reliable analysis, hence, requires simulations on several 
large lattices of the same structure. 
Due to the self-similar character of the underlying network,
the sizes of the lattices increase as a geometrical series.
We face the high C.P.U. time consuming problem in the simulations.
\end{enumerate}

The purpose of this paper is to provide a thorough study of the 3-state 
Potts model on the deterministic Sierpinski carpets of a given fractal 
dimension. 
The order of the transition has to be carefully checked before calculating 
the critical temperature and exponents. 
The paper is organized as follows: 
The model and the finite size scaling theory are briefly recalled
in Sec.2.1 and Sec.2.2. 
The simulation methods are described in Sec.2.3. 
The critical behavior of the $3$-state Potts model on the deterministic 
Sierpinski carpet is studied in Sec.3.

\section{Methods and theoretical background} 
\subsection{Sierpinski carpet $SC(3,1)$ and Potts model}
\qquad The lattice structures we deal with are constructed according to an 
iterative segmentation process: we start from a generating cell denoted 
$SC(3,1,1)$, which is a $3\times 3$ square lattice where the central square 
is removed.
The lattice $SC(3,1,k)$ associated with the $k^{th}$ iteration step
is constructed by replacing each occupied square of $SC(3,1,k-1)$ by the
generating cell while enlarging the side $3$ times. The size of the lattice $%
SC(3,1,k)$ is $L=3^{k}$ and the number of occupied sites $%
N_{L}^{occ}=(3^{2}-1^{2})^{k}$. The Hausdorff dimension $d_{f}$, defined in
such a manner that $N_{L}^{occ}=L^{d_{f}}$, is equal to $\ln (8)/\ln
(3)\approx 1.892789$. The Sierpinski carpet $SC(3,1,k)$ becomes a \textit{%
true} fractal in the mathematical sense when $k$ tends to infinity, and we
will denote it $SC(3,1)$.

The Hamiltonian of the $q$-state Potts model on $SC(3,1,k)$
reads : 
\begin{equation}
H=-J\sum_{\langle i,j\rangle }\delta (\sigma _{i},\sigma _{j}),
\end{equation}
where $\sigma _{i}$ and $\sigma _{j}$ stand for the spin states on the sites 
$i$ and $j$ and can take the integer values $1,2,\ldots ,q$. $\delta (\sigma
_{i},\sigma _{j})$ is the Kronecker $\delta $-function, and $J$ is the
coupling constant. The sum runs over the nearest-neighbor bonds on $SC(3,1,k)$.
The order parameter per site of the Potts model is defined as : 
\begin{equation}
m_{L}=\frac{q\rho _{L}-1}{q-1}
\end{equation}
where $\rho _{L}=\max (n_{1},n_{2},\ldots ,n_{q})/N_{L}^{occ}$ is the
largest density of spin state, since $n_{s}$ is the number of spins in state 
$s$. We notice that the values of $m_{L}$ lie in a range between $0$
and $1$. We will focus our attention on the $3$-state Potts model.

\subsection{Finite size scaling for second order and first order phase
transitions}
\qquad 
Finite size scaling theory has been first developed by Fisher 
{\it et al.} \cite{fisher72} in the case of the second order phase transitions.
According to Widom's homogeneity hypothesis, applying to a system with 
translational symmetry,
the singular part of the free energy per spin of the system 
under a change of the unit length from $1$ to $b$ 
is assumed to scale as 
\begin{equation}
f(t,h,L)=b^{-d} f(tb^{y_{t}},hb^{y_{h}},L/b)  \label{Fscal}
\end{equation}
where 
$d$ is the lattice dimension,
$L$ is the original lattice size,
$y_{t}$ and $y_{h}$ are eigenvalue exponents, 
$t=(T-T_{c})/T_{c}$ is the reduced temperature 
with $T_c$ representing the critical temperature on the infinite lattice,
and $h$ is the external field; 
$t$ and $h$ are supposed to be small, and $b$ smaller than the correlation 
length $\xi$.
This hypothesis (Eq.(\ref{Fscal})) has to be generalized in the case of 
fractals by restricting the choice of $b$ to a power of the size of the
generating cell of the Sierpinski lattice 
in order to keep invariant the structure of the renormalized lattice. 
Also, one has to replace the dimension $d$ in Eq.(\ref{Fscal})
by the Hausdorff dimension $d_f$ of the fractal; 
the factor $b^{-d_f}$ describes how the number of spins decreases under 
the change of the unit length.
Moreover, since the translational symmetry is lost on Sierpinski lattices,
the two point correlation function depends on the positions of the spins. 
Hence, the definition of the correlation length on the fractal lattices
has to be adapted.
A position-independent correlation function of one scalar variable $r$ 
can be defined by taking the average of the two point correlation function 
values over all the pairs of spins having direct distance equal to $r$.
The position-independent correlation function  behaves as 
$\exp(-r/\xi(T))/r^{d-2+\eta}$ in the critical region in the case of  
translational symmetry networks. 
We assume that it keeps a similar behavior on fractal lattices too.
For a second order phase transition system, $\xi(T)$ diverges at the critical 
point. 
We, therefore,  expect that $\xi(T)$ reaches the finite lattice size $L$ 
while $T$ is in the neighborhood of $T_c$. 
In this temperature region, we can take $b$ equal to the limiting value $L$
and it yields the zero-external field scaling equations for
the fractal lattice as: 
\begin{eqnarray}
C_{L}(T) &\sim &L^{\alpha /\nu }\,T\,\mathcal{C}(tL^{1/\nu }) \\
m_{L}(T) &\sim &L^{-\beta /\nu }\,\mathcal{M}(tL^{1/\nu }) \\
\chi _{L}(T) &\sim &L^{\gamma /\nu }\,\mathcal{X}(tL^{1/\nu }) \\
\phi _{L}(T) &\sim &L^{1/\nu }\,T^{2}\,\mathcal{P}(tL^{1/\nu }) \\
U_{L}(T) &\sim &T\,\mathcal{U}(tL^{1/\nu })
\end{eqnarray}
where $C_{L}(T)$ is the heat capacity per site, 
$m_{L}(T)$ the order parameter per site, 
$\chi _{L}(T)$ the zero field susceptibility of the order parameter per site, 
$\phi_{L}(T)\equiv\partial\ln\langle m_{L}(T)\rangle /\partial (k_{B}T)^{-1}$
usually called the first logarithmic derivative of the order parameter, and 
$U_{L}(T)\equiv 1-\langle m_{L}^{4}(T)\rangle /(3\langle m_{L}^{2}(T)\rangle
^{2})$ the Binder's cumulant.
The critical exponents $\alpha $, $\beta $, $\gamma $ and $\nu $ can be
written as a function of the two independent eigenvalue exponents $y_{t}$ 
and $y_{h}$ and the dimension $d_{f}$. 
$\alpha $ is equal to $2-d_{f}/y_{t}$, 
$\beta $ is equal to $(d_{f}-y_{h})/y_{t}$, 
$\gamma $ is equal to $ (2y_{h}-d_{f})/y_{t}$, 
and $\nu $ is equal to $1/y_{t}$. 
The finite-size shifting of the rounded singularity of a given thermodynamical
average $\kappa$ (for instance, $\phi $, $\chi $, and $C$)
from the critical point should follow the relation : 
\begin{equation}
T^{\kappa }(L)=T_{c}+g_{\kappa }L^{-1/\nu },  \label{Tscal}
\end{equation}
where  the $g_{\kappa }$'s are some constants. 
Moreover, the width of the associated rounded singularity should scale 
as $L^{-1/\nu }$.

For a system undergoing a first order phase transition, Fisher {\it et al.} 
\cite{fisher82} and Cardy {\it et al.} \cite{cardy83} propose that the
relevant eigenvalue exponents in Eq.(\ref{Fscal}) should reach their
allowed limiting values, said $y_{t}=d$ and $y_{h}=d$, in the
phenomenological studies. 
In the case of the Sierpinski lattice, $y_{t}$ and $y_{h}$ should take the
value of the Hausdorff dimension $d_f$. 
The thermodynamical averages on the fractal lattice, therefore, scale as: 
\begin{eqnarray}
C_{L}(T) &\sim &L^{d_{f}}\,T\,\mathcal{C}(tL^{d_{f}}) \\
\chi _{L}(T) &\sim &L^{d_{f}}\,\mathcal{X}(tL^{d_{f}}) \\
\phi _{L}(T) &\sim &L^{d_{f}}\,T^{2}\,\mathcal{P}(tL^{d_{f}})
\end{eqnarray}
Moreover, they are characterized by $\delta$-function like singularities as the
lattice size $L$ increases ; as a result, the areas under the curves
displaying a given average as a function of the temperature should be
independent of $L$. 
Also, it is well known that the probability distribution of the energy is 
double peaked in the vicinity of the transition temperature \cite{kerler93}. 

\subsection{Simulation process}
\qquad 
We implement the Wolff \cite{wolff88} and Swendsen-Wang \cite{sw87} 
Monte Carlo cluster algorithms;
both of them have been shown to largely overcome the critical slowing down. 
In order to save simulation time, we used alternatively the two algorithms 
in our studies. 
Let us describe the simulation process in a more precise way. 
An initial configuration is built up by randomly assigning one of the three
spin states to each occupied site of the lattice $SC(3,1,k)$. Periodic
boundary conditions will be used. A simulation temperature $T_{0}$ is set
and enough Wolff or Swendsen-Wang Monte Carlo steps are performed in order
to ensure that the configuration has been thermalized. The energy per site 
$\epsilon _{L}(T_{0})$ and the order parameter per site $m_{L}(T_{0})$ are
then calculated for every Monte Carlo step. 
One million steps are collected to be a sample and 
the above procedure is started again at least ten times. 
The histogram method \cite{ferr89}, applied within each sample, 
allows the thermodynamical averages of a physical quantity $g$ to 
be calculated over a range $\Delta T$ around 
$T_{0}$; we denote it $\langle g(T_{0})\rangle _{T}$. 
The reliability of the calculated averages over the temperature interval
must be carefully checked on. 
$\phi_{L}(T)$,  $\chi _{L}(T)$,  $C_{L}(T)$, and $U_{L}(T)$ 
are then obtained according to the following relations: 
\begin{eqnarray}
\phi _{L}(T) &=&N_{L}^{occ}\left( \langle \epsilon _{L}(T_{0})\rangle _{T}-%
\frac{\langle \epsilon _{L}(T_{0})\,m_{L}(T_{0})\rangle _{T}}{\langle
m_{L}(T_{0})\rangle _{T}}\right)  \\
\chi _{L}(T) &=&\frac{N_{L}^{occ}(\langle m_{L}^{2}(T_{0})\rangle
_{T}-\langle m_{L}(T_{0})\rangle _{T}^{2})}{k_{B}T} \\
C_{L}(T) &=&\frac{N_{L}^{occ}(\langle \epsilon _{L}^{2}(T_{0})\rangle
_{T}-\langle \epsilon _{L}(T_{0})\rangle _{T}^{2})}{k_{B}T^{2}} \\
U_{L}(T) &=&1-\frac{\langle m_{L}^{4}(T_{0})\rangle _{T}}{3\langle
m_{L}^{2}(T_{0})\rangle _{T}^{2}}.
\end{eqnarray}
The statistical errors associated with the thermodynamical averages are
estimated by calculating the standard deviations from the ten samples.

\section{Results and discussions} 
\subsection{Studies of rounded singularities of $\protect\chi_{L}(T)$ and 
$\phi_{L}(T)$}
\qquad First, we study the order of the phase transition of the $3$-state
Potts model in the case of the fractal $SC(3,1)$. The behavior of the
finite-size rounded singularities of $\chi_{L}(T)$ calculated from our
simulations are shown in Fig.$1$ for $L=27,\ 81,\ 243,\ 729,$ and $2187$.
The maximum values $\chi_{L}^{max}$ and their associated temperatures 
$T^{\chi}(L)$ are reported in Table 1. The finite-size shifts of 
$\chi_{L}^{max}$ are clearly observed; the Lorentz distribution is a better
candidate to fit the rounded singularity than the Gaussian one: 
\begin{equation}
\chi _{L}(T)=\frac{2A}{\pi }\left[ \frac{W}{4(T-T^{\chi }(L))^{2}+W^{2}}
\right]  \label{lorentz}
\end{equation}
The width of the Lorentz distribution is $W$ and the area under the
distribution curve is $A$. The two-parameter fits $(W,\,A)$ of the rounded
singularities are reported in Table 2. It appears clearly that the area 
$A$ under the Lorentz distribution increases as $L$ increases; as a result
of this, we can conclude that the singularity is not $\delta$-function like
as $L$ tends to infinity. 
We have the first evidence that the phase transition is not a first order one. 
According to the finite size scaling for second order phase transitions, 
the width $W$ should scale as $L^{-1/\nu }$ and the area $A$ as 
$L^{\gamma/\nu -1/\nu}$. 
The log-log plots of $W$ and $A$ versus $L$ do not appear straight lines 
over the whole range from $L=27$ to $L=2187$; 
they exhibit a slight curvature, which can be interpreted as a
contribution of scaling corrections to the behavior of $W$ and $A$. 
The slopes calculated from two consecutive points from $L=27$ to $L=2187$ are 
$-0.786(17)$, $-0.701(11)$, $-0.6243(60)$, $-0.5524(35)$ for $W$ and 
$0.897(12)$, $0.9867(81)$, $1.0635(40)$, $1.1634(25)$ for $A$. 
The effect of the scaling corrections in the case of fractal lattices has 
been pointed out and discussed in the framework of the Ising model by 
Monceau {\it et al.} \cite{monceau01} and Carmona {\it et al.} \cite{carmona98}.
Instead of a power law, $W$ is expected to behave as 
$L^{-1/\nu }(1+\varphi_{W}(L))$
where the scaling corrections $\varphi_{w}(L)$ tend to zero as $L$ tends to
infinity; usually, $\varphi_{W}(L)$ is developed as 
$\varphi_{W}(L)={\cal A}_{W}L^{-\omega }$, 
where ${\cal A}_{W}$ is the amplitude of the corrections and 
$\omega$ an additional exponent. 
A similar behavior is expected for $A$. 
There is no hope to calculate the scaling corrections 
in a reliable way from our simulations. 
However, the slopes calculated above enable to estimate an upper bound for 
$1/\nu $ to be $0.5524(35)$ and a lower bound for $\gamma /\nu -1/\nu $ 
to be $1.1634(25)$.

We now pay attention to the maxima $\chi_{L}^{max}$ of $\chi_{L}(T),$ and 
$\phi _{L}^{max}$ of $\phi_{L}(T)$, reported in Table 1. The log-log plot
of $\chi_{L}^{max}$ shows a straight line. The power law fits are satisfied
with reliability coefficients equal to $1.0000$; 
they yield $\gamma /\nu=1.6948(37)$ with a four sizes fit from $L=81$ to 
$L=2187$ 
and $\gamma /\nu =1.7013(28)$  with a three sizes fit from $L=243$ to $L=2187$. 
The scaling corrections turn out to be very weak in the case of the maxima of 
the susceptibility, as already shown in the case of the Ising model. 
The best estimate of $\gamma /\nu $ is $1.7013(28)$. 
Finite size scaling theory states that the slope of the line
represents the exponent $\gamma /\nu $ if the phase transition is second
order and that it should be the Hausdorff dimension $d_{f}$ if the
transition is first order. The stability of the fits involving 
$\chi_{L}^{max}$ enable to conclude that $\gamma /\nu $ is significantly
different from $d_{f}=1.892789$ and, therefore, consistent with a second order 
phase transition as we have shown at the beginning of this paragraph.
On the other hand, $\phi_{L}^{max}$ suffers from large scaling corrections
since the slopes of two successive points of $\phi_{L}^{max}$ in a log-log 
plot are measured as $0.758(20)$, $0.669(11)$, $0.613(13)$, and $0.533(11)$. 
These large scaling corrections may be due to the deviation of 
the average number of nearest neighbors on the finite size fractal from 
that on the infinite fractal \cite{monceau01}
because the critical temperature is directly related to the average number
of nearest neighbors in the mean filed theory and 
the eignvalue exponent $y_t=1/\nu$ describes the scaling behavior of
the reduced temperature.  
The behavior of $\phi _{L}^{max}$ provides an upper bound for $1/\nu$, 
which is $1/\nu =y_{t}\leq 0.533(11)$.
Combining $\gamma/\nu =1.7013(28)$ with $1/\nu\leq 0.533(11)$, we
get a lower bound for $(\gamma/\nu -1/\nu)$, which is $1.168(11)$. 
These bounds are both very close to the ones we got from the fits 
of $W$ and $A$.
The eigenvalue exponent $y_{h}$ can be calculated from $(\gamma/\nu +d_{f})/2$
and is equal to $1.7970(15)$.

The comparison between the present value of $\gamma/\nu$ and 
the one associated with the Ising model for the same fractal structure 
$(\gamma /\nu= 1.732(4))$ \cite{monceau01} 
enables to distinguish the Ising and the 3-state Potts model
universality classes. 
We mention that this comparison makes sense only if the fractal structures 
are exactly the same because the universality is said to be weak 
\cite{gefen80, hsiao00, wu87}. 

\subsection{Estimation of the critical temperature $T_{c}$}
\qquad The infinite lattice critical temperature $T_{c}$ can be calculated
from the finite-size shifts of the rounded singularities\ provided that $\nu$
is known; for a given peaked thermodynamical average $\kappa $, the
equation $(\ref{Tscal})$ yields:
\begin{equation}
T_{c}=\frac{T^{\kappa }(\ell L)-\ell^{-1/\nu }\,T^{\kappa }(L)}
{1-\ell ^{-1/\nu }}  \label{TCserie}
\end{equation}
where $\ell =3$ is the size of the generating cell. 
In this form, the value of $T_{c}$ is deduced from two consecutive iteration 
steps of the fractal structure. 
An upper bound for $1/\nu $ will yield an upper bound for $T_{c}$ 
since $T^{\chi }(L)$ and $T^{\phi }(L)$ decrease as L increases.
Taking $1/\nu =0.533(11)$, the values of $T^{\chi }(L)$ from the two
highest iteration steps provide the first estimate for the upper bound of 
$T_{c}$ equal to $0.67324(1264)$. The values of  $T^{\phi}(L)$ from the
two highest iteration steps yield $0.67357(1456)$ as a second upper bound 
for $T_{c}$. 
The errors in the above bounds mainly come from the error propagation of 
$1/\nu$. 
We can conclude that $T_{c}<0.673(12)$.

An alternative way to measure $T_{c}$ without knowing $\nu $ is to study the
Binder's cumulant crossing points \cite{binder81}. 
Fig.$2$ shows that the cumulant curves for different lattice sizes do not 
intersect at a fixed point. 
We call $T_{k-1,k}$ the temperature of the cumulant intersection
point between the iteration steps $(k-1)$ and $k$. 
The measured values are: $T_{2,3}=0.68461(54)$, $T_{3,4}=0.67923(26)$, 
$T_{4,5}=0.67594(20)$, $T_{5,6}=0.673916(69)$, and $T_{6,7}=0.672866(50)$. 
The difference $\Delta T_{k}=(T_{k-1,k}-T_{k,k+1})$ behaves basically 
as a geometrical series as a function of $k$.
If we write $\Delta T_{k}=ar_{3}r_{4}\cdots r_{k}$ 
where $r_{k}=\Delta T_{k}/\Delta T_{k-1}$, $k\geq 4$, 
and $ar_{3}=\Delta T_{3}$, 
we will find $ 1 > r_{4} \gtrsim r_{5} \gtrsim r_{6} \gtrsim \cdots$. 
This enables us to claim that the series $\{T_{k,k+1}\}$ converges as $k$ 
tends to infinity.
The convergent value calculated from the three intersection points $T_{4,5}$, 
$T_{5,6}$, and $T_{6,7}$ gives a lower bound for the critical temperature: 
$0.671737$. We finally conclude that $0.671737<T_{c}<T_{6,7}=0.672866$.
This range is in agreement with the upper bound for $T_{c}$ extracted from
Eq.(\ref{Tscal}).

\subsection{Simulation near $T_c$}
\qquad 
We firstly check the occurence of a single peak in the energy probability
distributions, which confirms again that the transition is second order.
A typical energy probability distribution is shown in Fig.$3$. 
We then study the behaviors of $\phi _{L}(T)$, $\chi _{L}(T)$, and $m_{L}(T)$ 
within the previously estimated temperature range for $T_{c}$. 
The following three temperatures will be considered: $0.671737$, $0.672866$, 
and $0.673241$. 
The first two temperatures are the lower and the upper bounds
extracted from the Binder's cumulant crossing and the last one is the upper
bound calculated from Eq.$(\ref{TCserie})$, assuming that $1/\nu =0.533$ and
using the positions of the peaks of $\chi_{L}$ at $L=729$ and $L=2187$. 
The log-log plots of $m_{L}(T)$, $\chi_{L}(T)$, and $\phi_{L}(T)$ at the 
three considered temperatures are shown in Fig.$4$. 
$m_{L}$ shows a straight line in the log-log plot, at $T=0.671737$, over a
range of lattice sizes covering several orders of magnitude, with a
slope, measured from $L=243$, $L=729$, and $L=2187$, equal to $-0.0748(1)$.
When the temperature increases towards the upper bound, $m_{L}(T)$ becomes
more and more concave while $\chi_{L}(T)$ is more and more convex. 
The asymptotic slopes for $m_{L}(T)$ obtained from the measurements beween 
the largest sizes $L=729$ and $L=2187$ are upper bounded 
at $T=0.672866$ and $T=0.673241$, respectively, by 
$-0.0876(2)$ and $-0.0932(2)$; 
and the asymptotic slopes for $\chi _{L}(T)$ at $T=0.671737$, $T=0.672866$, 
and $T=0.673241$ are lower bounded by $1.60(1)$, $1.73(1)$, and $1.78(1)$.
On the other hand, the curvature of $\phi_{L}(T)$ in a log-log plot varies 
in the opposite direction, that is, the curve is less convex at the higher 
temperature bound ($T=0.673241$) than at the lower one ($T=0.671737$). 
The asymptotic slopes for $\phi_{L}(T)$ at $ T=0.671737$, $T=0.672866$, 
and $T=0.673241$ are upper bounded by $0.466(13)$, $0.521(10)$, $0.552(11)$. 
The scaling corrections, depending on the physical quantities, 
turn out to affect strongly the critical behavior of the 3-state Potts model
on the fractal structure even when the sizes are large. 
The estimation of the ratio of critical exponents $\beta/\nu$, 
$\gamma/\nu$, and $1/\nu$ from the behavior of $m_{L}$, $\chi_{L}$,
and $\phi_{L}$ in the critical region becomes a very difficult task: 
firstly, the value of the critical temperature cannot be provided precisely; 
secondly, even if we expect scaling corrections to tend to zero 
as $L$ tends towards infinity (whatever thermodynamic quantity is studied), 
they should be included in the calculation of our exponents; 
as already explained in Sec.3.1, there is no hope to extract 
scaling corrections from our data.  
We point out that the scaling corrections in the critical region  
are stronger than that for the case of Ising model.
In the case of Ising model, the slope of $\chi_{L}(T)$ extracted at the 
expected critical temperature is consistent with the value of $\gamma/\nu$ 
extracted from $\chi_{L}^{max}$ \cite{hsiao00, monceau01}. 

\subsection{Discussions}
\qquad
The consistency of our full set of results can be achieved in the following 
way: 
The behavior of the Binder cumulant shows that $T_{c}$ should be greater 
than $0.671737$ and smaller than $0.682866$. 
From the behavior of $\chi_{L}(T)$, the asymptotic slopes in this temperature
region are lower bounded by 1.60(1), consistent with the value 
$\gamma/\nu=1.7013(29)$ extracted from $\chi_{L}^{max}$. 
The asymptotic slopes for $m_{L}(T)$ in the region are always smaller than 
-0.0748(1) and, thus, yield $\beta/\nu >0.0748(1)$,
consistent with the requirement of the hyperscaling relation, said
$\beta/\nu=(d_{f}-\gamma/\nu)/2=0.0957(21)$ with $\gamma/\nu$ taking
the value 1.7013(29).
Similarly, $1/\nu <0.521(11)$ can be found from the behavior of
$\phi_{L}(T)$ in this temperature region, consistent with the results
in the studies of $\phi_{L}^{max}$. 
We, therefore, give the best value and bounds of the exponents and 
the critical temperature that we find: 
$\gamma/\nu =1.7013(29)$, $\beta/\nu>0.0748(1)$, $1/\nu<0.521(11)$, and 
$0.671737<T_{c}<0.672866$.

The critical temperature and the exponents of the 3-state Potts model 
on a two dimensional regular system are known to take the following values
\cite{baxter82}:
$T_c= 1/\ln(1+\sqrt{3}) \simeq 0.995$, $1/\nu=6/5$, $\beta/\nu=2/15$,
and $\gamma/\nu=26/15$.  
We find that the values of $T_c$, $1/\nu$, $\beta/\nu$ and $\gamma/\nu$ 
for the 3-state Potts model are smaller in the case of the Sierpinski
carpet we investigated than in the case of the two dimensional regular system.
A similar situation occurs in the case of the Ising model on Sierpinski 
carpets \cite{monceau01} where, regardless of the structure of the 
fractals, those values decrease as the Hausdorff dimension decreases. 

Discrepancies are observed between our simulation results and Bin's predictions.
Bin \cite{bin86} found  
by applying the method of Migdal-Kadanoff bond-moving renormalization,
that $\exp(-K)=0.746$ (which is equivalent to $T_{c}=3.4126$) and 
$y_{K}(=y_t)=0.619$ for the 3-state Potts model on the same Sierpinski
carpet we studied in this paper. 
There are mainly two reasons for the occurence of discrepancies:
firstly, the Migdal-Kadanoff bond-moving renormalization is an 
approximate method;
secondly, in Bin's case the spins were located at the vertices of the occupied 
squares of the Sierpinski carpet while they are located at the centers of the 
occupied squares in the present work;
it yields a different universality class because the duality is no more held
on fractal lattices. 

In conclusion, we have carefully performed the first Monte Carlo studies for 
the $3$-state Potts model on the Sierpinski carpet with $d_f \simeq 1.8928$.
Although the system exhibits strong scaling corrections,
the consistency of our results demonstrates the validity of the
finite size scaling theory for the $q$-state Potts model on these fractal 
lattices. 
Nevertheless, scaling corrections tend to increase while $q$ is passing 
from $2$ (Ising model) to $3$. 
We, hence, expect the raise of these difficulties in the studies of 
the phase transition while $q$ goes larger. 
In the coming future, we will be able to perform in a reliable way
the studies for the Potts model with larger $q$ on fractal dimensions higher
than 2. 

\section*{Acknowledgements}
A part of numerical simulations has been carried out in the national center
of computational resources IDRIS, supported by the CNRS (project number $%
991186$). We acknowledge the scientific committee and the staff of the
center. We are also grateful to the research computational center (CCR) of
the university of Paris VII -- Denis~Diderot, where the rest of the simulations
have been done.

\clearpage  

\section*{Figure captions} 
\begin{enumerate}
\item[Fig.1] Rounded singularities of the susceptibility on lattice $SC(3,1,k)$
where $k=3$, 4, 5, 6, and 7. Each segment represents the reliabe temperature
range of the histogram method where the simulation temperature is
located in its center. The fitting by Lorentz distribution for the
susceptibility on $SC(3,1,7)$ is plot in order to clearly indicate the
location of the rounded singularity.

\item[Fig.2]  Binder's cumulant crossings. 

\item[Fig.3]  Probability distribution of energy per site on $SC(3,1,7)$.
It's made of 10 million Monte Carlo data and seperated by 500 histogram
classes. The simulation temperature is taken at 0.672845. 

\item[Fig.4]  Log-log plot of $m_L$, $\chi_L$, and $\phi_L$ at $T=0.671737$
(open square), $0.672866$ (open circle), and $0.673241$ (open triangle) 
on the Sierpinski carpet $SC(3,1,k)$. 
\end{enumerate}

\clearpage  
\begin{table}
\caption{Maxima values and associated peak positions of $\chi_{L}$ and 
$\phi_{L}$  on the Sierpinski carpet $SC(3,1,k)$ of size $L=3^k$}
\begin{tabular}{|c||c|c|c|c|c|}
\hline
& $L=27$ & $L=81$ & $L=243$ & $L=729$ & $L=2187$ \\ \hline
$\chi _{L}^{max}$ & 30.53(12) & 195.58(59) & 1239.4(4.4) & 7984(19) & 
51969(105) \\ 
$T^{\chi }(L)$ & 0.73800(22) & 0.70448(12) & 0.689525(65) & 0.682115(16) & 
0.678182(11) \\ \hline
$\phi _{L}^{max}$ & 8.85(18) & 20.35(13) & 42.46(45) & 83.23(74) & 
149.50(1.23) \\ 
$T^{\phi }(L)$ & 0.7510(17) & 0.70995(38) & 0.69221(22) & 0.68357(16) & 
0.679139(23) \\ \hline
\end{tabular}
\end{table}

\begin{table}
\caption{Fit of the Lorentz distribution (Eq.(\ref{lorentz})) of the rounded 
singularity of $\chi_L$ on the Sierpinski carpet $SC(3,1,k)$ of size $L=3^k$.}
\begin{tabular}{|c||c|c|c|c|c|}
\hline
& $L=27$ & $L=81$ & $L=243$ & $L=729$ & $L=2187$ \\ \hline
$W$ & 0.0700(11) & 0.02949(31) & 0.01366(8) & 0.00688(2) & 0.00375(1) \\ 
$A$ & 3.444(38) & 9.230(74) & 27.29(11) & 87.78(18) & 315.13(57) \\ \hline
\end{tabular}
\end{table}

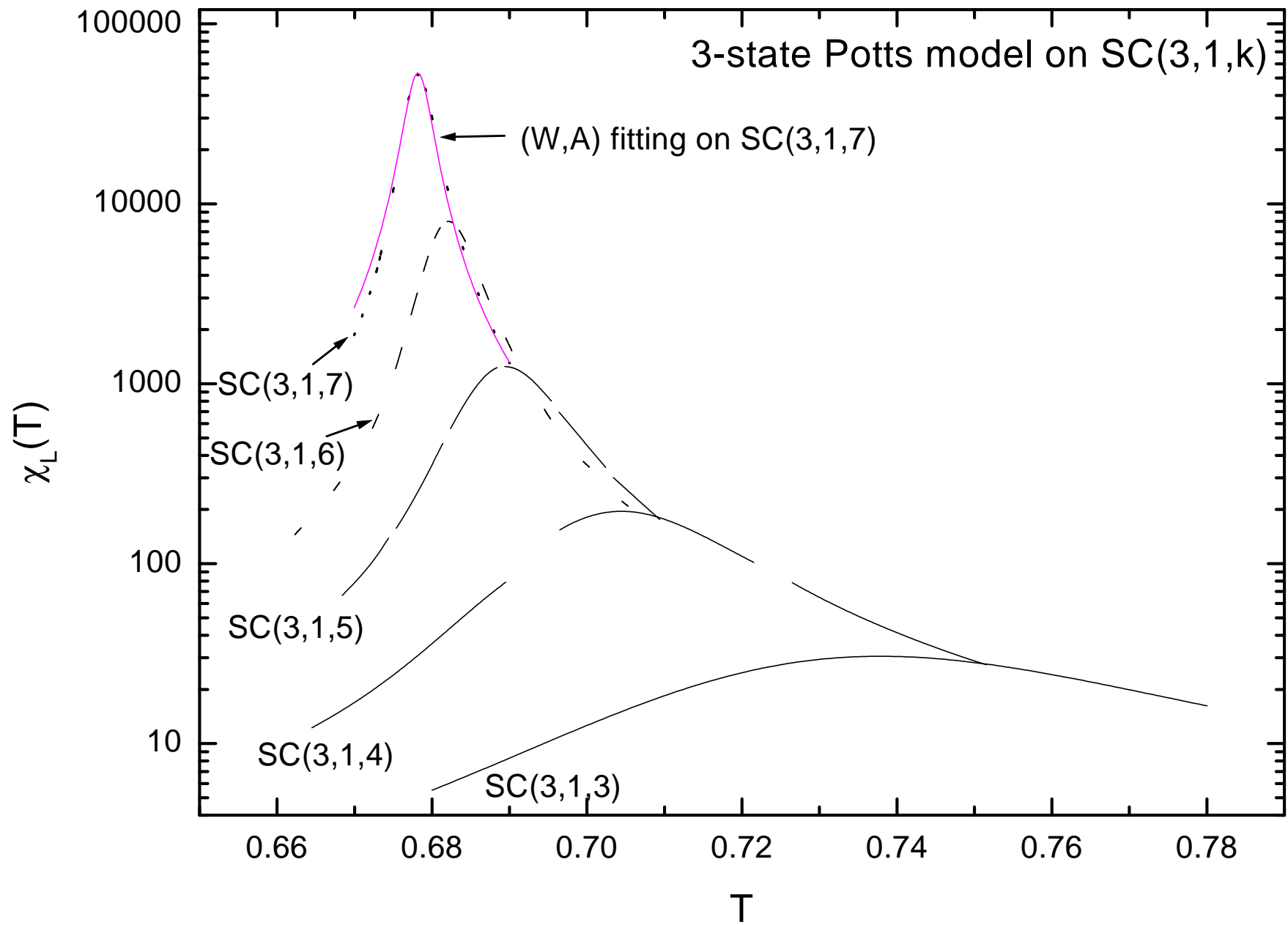
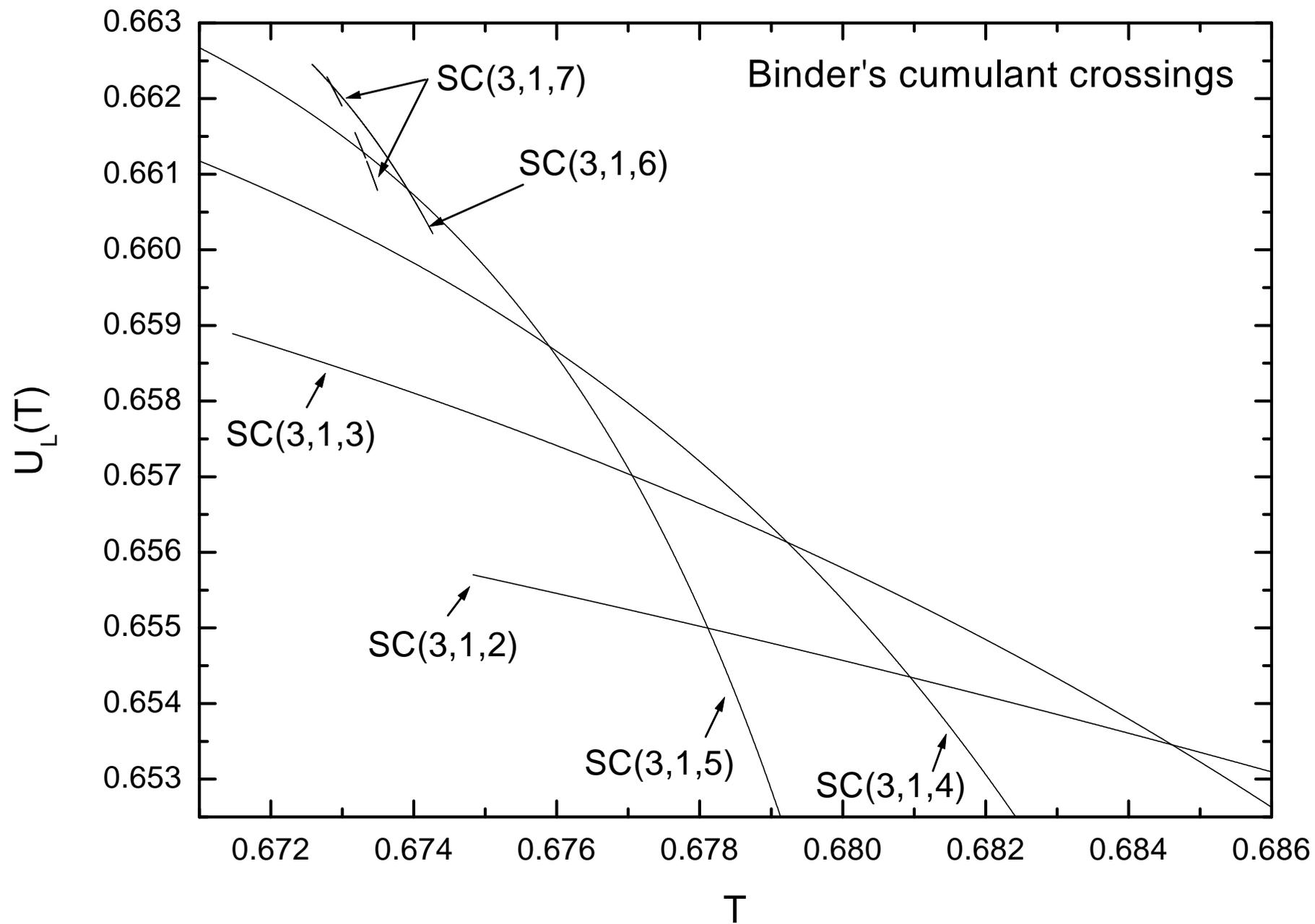
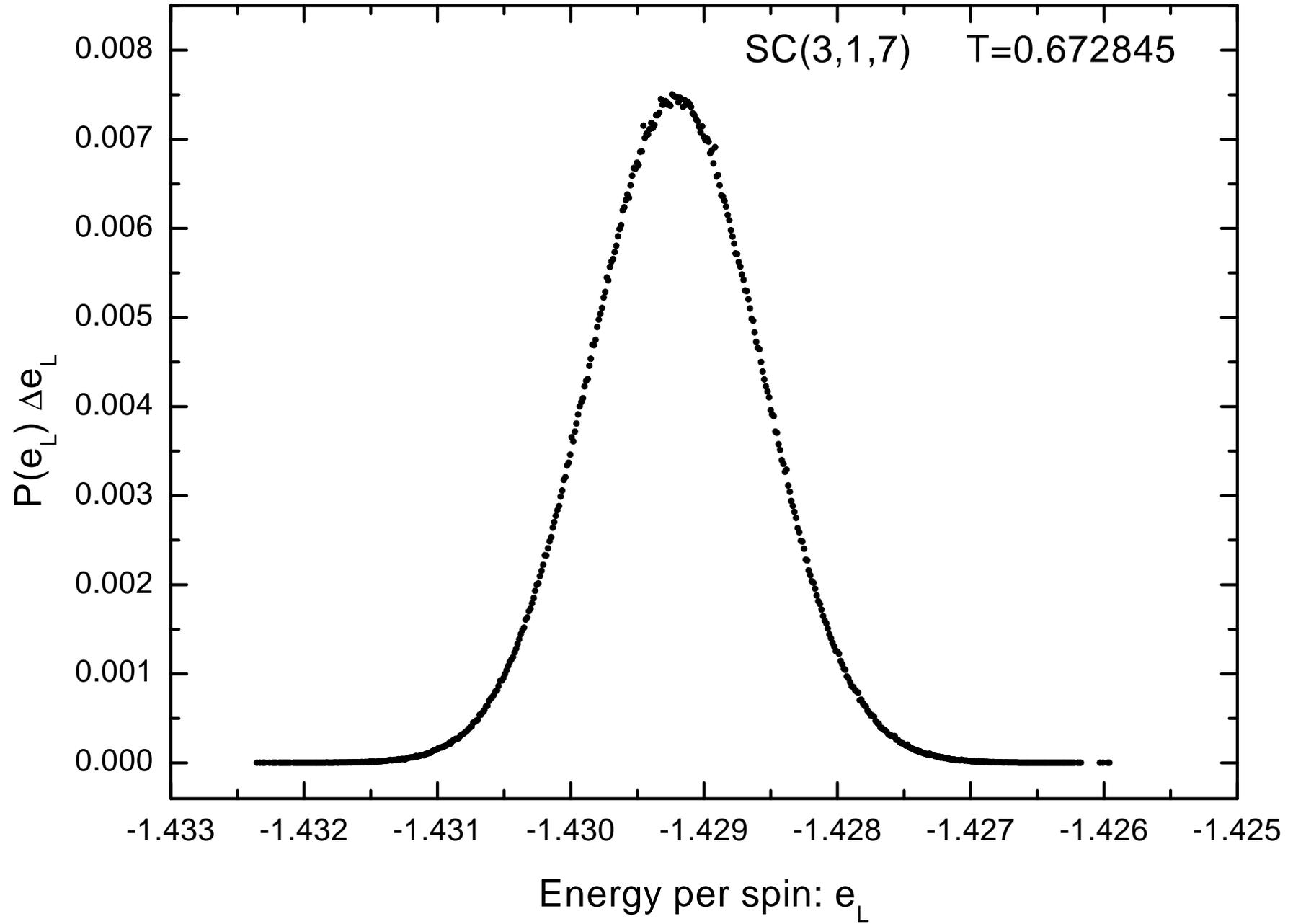
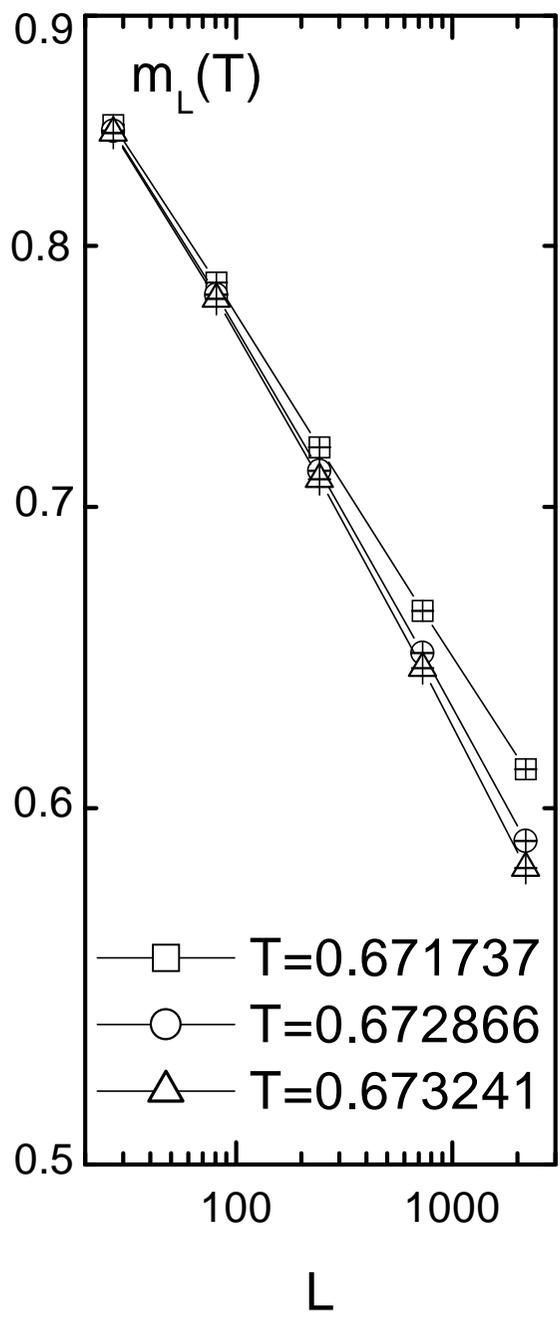
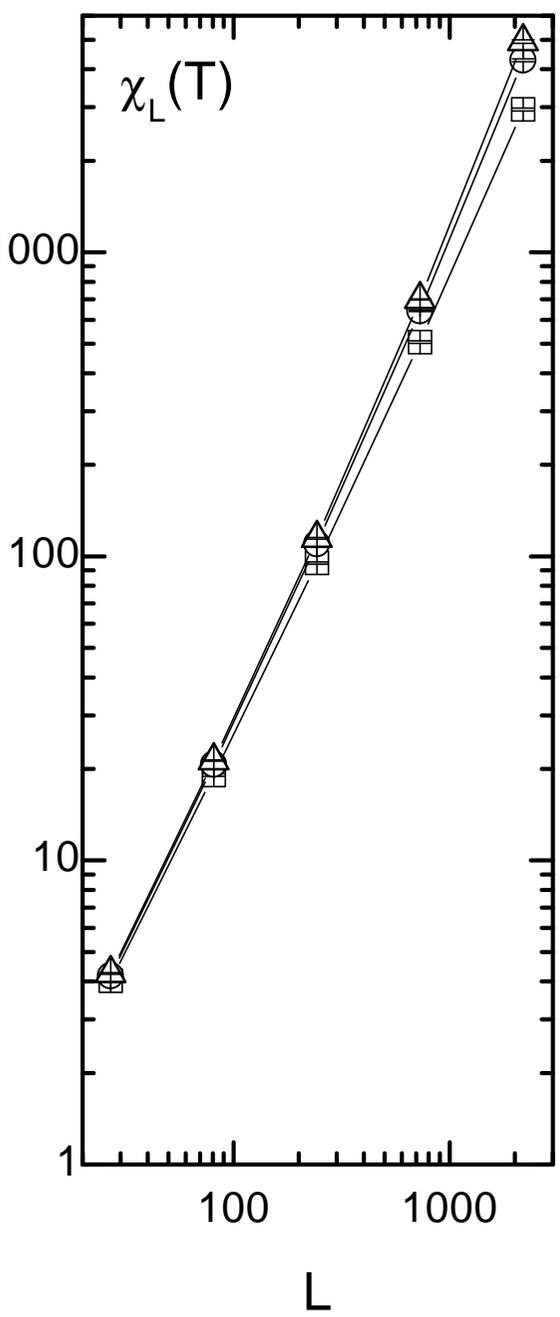
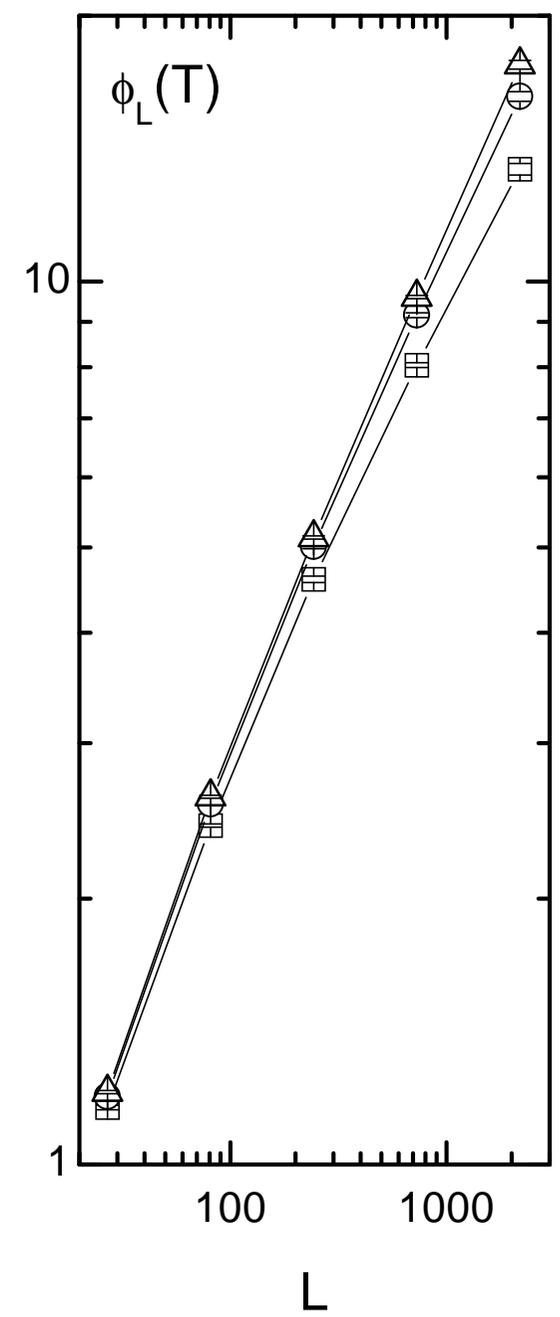
\end{document}